\newcommand{\nar}{New Astron. Rev.}

 \documentclass[final,authoryear,5p,times,twocolumn]{elsarticle}

\usepackage[super]{nth}

\usepackage{graphicx}
  \graphicspath{{epsfigs/}}
 \usepackage{amsbsy}

\usepackage{amssymb}
\journal{Icarus}

\usepackage{natbib}
\bibpunct{(}{)}{;}{a}{}{,}
\usepackage{hyperref}
\hypersetup{
    bookmarks=true,         % show bookmarks bar?
    unicode=false,          % non-Latin characters in Acrobat's bookmarks
    pdftoolbar=true,        % show Acrobat's toolbar?
    pdfmenubar=true,        % show Acrobat's menu?
    pdffitwindow=false,     % page fit to window when opened
    pdfsubject={Planetary Science}, 
    pdftitle={The small binary asteroid (939) Isberga}, 
    pdfkeywords={},         % list of keywords
    pdfnewwindow=true,      % links in new window
    colorlinks=true,        % false: boxed links; true: colored links
    linkcolor=gray,         % color of internal links
    citecolor=blue,         % color of links to bibliography
    filecolor=gray,         % color of file links
    urlcolor=gray           % color of external links
}

\usepackage{multirow}

\newcommand{\eg}{e.g.}
\newcommand{\ie}{i.e.}

\newcommand{\degr}{\ensuremath{^\circ}}
\newcommand{\arcsec}{\mbox{\ensuremath{^{\prime\prime}}}}

\usepackage{ulem}

\newcommand{\ben}[1]{#1}

\begin{document}

\begin{frontmatter}

%% Title, authors and addresses

%% use the tnoteref command within \title for footnotes;
%% use the tnotetext command for the associated footnote;
%% use the fnref command within \author or \address for footnotes;
%% use the fntext command for the associated footnote;
%% use the corref command within \author for corresponding author footnotes;
%% use the cortext command for the associated footnote;
%% use the ead command for the email address,
%% and the form \ead[url] for the home page:
%%
%% \title{Title\tnoteref{label1}}
%% \tnotetext[label1]{}
%% \author{Name\corref{cor1}\fnref{label2}}
%% \ead{email address}
%% \ead[url]{home page}
%% \fntext[label2]{}
%% \cortext[cor1]{}
%% \address{Address\fnref{label3}}
%% \fntext[label3]{}

\title{The small binary asteroid (939) Isberga\tnoteref{088.C-0770}}
\tnotetext[088.C-0770]{Based on observations made with ESO telescopes at the
  La Silla Paranal Observatory under programme ID:
  \href{http://archive.eso.org/wdb/wdb/eso/sched_rep_arc/query?088.C-0770}{088.C-0770}} 

%% use optional labels to link authors explicitly to addresses:
%% \author[label1,label2]{<author name>}
%% \address[label1]{<address>}
%% \address[label2]{<address>}

\author[imcce,esac]{B.~Carry}
  \ead{bcarry@imcce.fr}
\author[mpa,grenoble]{A.~Matter}
\author[prague]{P.~Scheirich}
\author[prague]{P.~Pravec}

\author[molnar]{L.~Molnar}     %    LC
\author[dlr]{S.~Mottola}       %    LC
\author[carbo]{A.~Carbognani}  %    LC
\author[liege]{E.~Jehin}       %    LC
\author[poznan]{A.~Marciniak } %    LC

\author[mit]{R.~P.~Binzel}
\author[mit,harvard]{F.~E.~DeMeo}
\author[imcce]{M.~Birlan}

\author[nice]{M.~Delbo}
\author[cdr,aude]{E.~Barbotin}
\author[cdr,geneve]{R.~Behrend}
\author[cdr,aude]{M.~Bonnardeau}
\author[imcce]{F.~Colas}
\author[cala]{P.~Farissier}
\author[bardon,t60]{M.~Fauvaud}
\author[bardon,t60]{S.~Fauvaud}
\author[cala]{C.~Gillier}
\author[liege]{M.~Gillon}
\author[dlr]{S.~Hellmich}
\author[poznan]{R.~Hirsch}
\author[cdr]{A.~Leroy}
\author[liege]{J.~Manfroid}
\author[cdr]{J.~Montier}
\author[cdr]{E.~Morelle}
\author[t60]{F.~Richard}
\author[poznan]{K.~Sobkowiak}
\author[cdr]{J.~Strajnic}
\author[imcce]{F.~Vachier}

\address[imcce]{\ben{IMCCE, Observatoire de Paris, UPMC Paris-06, Universit{\'e}
Lille1,} UMR8028 CNRS, 77 av. Denfert Rochereau, 75014 Paris, France}
\address[esac]{European Space Astronomy Centre, ESA, P.O. Box 78, 28691 Villanueva de la Ca\~{n}ada, Madrid, Spain}
\address[mpa]{Max Planck institut f{\"u}r Radioastronomie, auf dem H{\"u}gel, 69, 53121 Bonn, Germany}
\address[grenoble]{UJF-Grenoble 1 / CNRS-INSU, Institut de Plan{\'e}tologie et d'Astrophysique de Grenoble (IPAG) UMR 5274, Grenoble, F-38041, France}
\address[prague]{Astronomical Institute, Academy of Sciences of the Czech
  Republic, Fri{\v c}ova 298, CZ-25165 Ond{\v r}ejov, Czech Republic}
\address[molnar]{Department of Physics and Astronomy, Calvin College, 3201 Burton SE, Grand Rapids, MI 49546, USA}
\address[dlr]{Deutsches Zentrum f{\"u}r Luft- und Raumfahrt (DLR), 12489 Berlin, Germany}
\address[nice]{UNS-CNRS-Observatoire de la C{\^o}te d’Azur, Laboratoire Lagrange, BP 4229 06304 Nice cedex 04, France}
\address[mit]{Department of Earth, Atmospheric and Planetary Sciences, MIT, 77 Massachusetts Avenue, Cambridge, MA, 02139, USA}
\address[harvard]{Harvard-Smithsonian Center for Astrophysics, 60 Garden Street, MS-16, Cambridge, Massachusetts 02138, USA}
\address[liege]{Institut d’Astrophysique et de G{\'e}ophysique, Universit{\'e}e de Li{\`e}ge, All{\'e}e du 6 ao{\^u}t 17, B-4000 Li{\`e}ge, Belgium}
\address[poznan]{Astronomical Observatory Institute, Faculty of Physics,
  A. Mickiewicz University, S\l oneczna 36, 60-286, Pozna{\'n}, Poland}
\address[carbo]{Astronomical Observatory of the Autonomous Region of the Aosta Valley, Loc. Lignan 39, 11020, Nus (Aosta), Italy}
\address[geneve]{Geneva Observatory, CH-1290 Sauverny, Switzerland}
\address[cdr]{CdR \& CdL Group: Lightcurves of Minor Planets and Variable Stars, Switzerland}
\address[aude]{Association des Utilisateurs de D{\'e}tecteurs {\'E}lectroniques (AUDE), France}
\address[cala]{Club d'Astronomie de Lyon Amp{\`e}re (CALA), Place de la Nation, 69120 Vaulx-en-Velin, France}
\address[bardon]{Observatoire du Bois de Bardon, 16110 Taponnat, France}
\address[t60]{Association T60, 14 avenue Edouard Belin, 31400 Toulouse, France}

\begin{abstract}
%-context
  \ben{In understanding} the composition and internal structure of
  asteroids, their density is perhaps the most diagnostic quantity.
%-aim
  We aim here to characterize the surface composition, mutual orbit,
  size, mass, and density of the small \ben{main-belt} binary asteroid (939) Isberga.
%-method
  For that, we conduct a suite of multi-technique observations, including
  optical lightcurves over many epochs, near-infrared spectroscopy,
  and interferometry in the thermal infrared.
  We develop a simple geometric model of binary systems to analyze the
  interferometric data in combination with the results of the
  lightcurve modeling.
%-results
  From spectroscopy, we classify Ibserga as a Sq-type asteroid,
  \ben{consistent with the albedo of
  0.14$^{+0.09}_{-0.06}$ (all uncertainties are reported
  as 3-$\sigma$ range) we determine 
  (average albedo of S-types is 0.197\,$\pm$\,0.153, see Pravec et
  al., 2012, Icarus 221, 365-387).}
  Lightcurve analysis reveals that
  the mutual orbit has a period of 26.6304\,$\pm$\,0.0001\,h,
  is close to circular (eccentricity lower than 0.1), 
  and has pole coordinates \ben{within 7\degr~of (225\degr,+86\degr)} in
  Ecliptic J2000, implying a low obliquity of $1.5^{+6.0}_{-1.5}$ degree.
  The combined analysis of lightcurves and interferometric data allows
  us to determine the dimension of the system and 
  we find volume-equivalent diameters of
  \ben{12.4$^{+2.5}_{-1.2}$\,km and 
        3.6$^{+0.7}_{-0.3}$\,km} for Isberga and its satellite, circling each
  other on a 33\,km wide orbit.
  \ben{Their density is assumed equal and found to be 
  $2.91^{+1.72}_{-2.01}$ g.cm$^{-3}$, lower than that of the
  associated ordinary chondrite meteorites, suggesting the presence of
  some macroporosity, but typical of S-types of
  the same size range (Carry, 2012, P\&SS 73, 98--118).}
%-discussion
  The present study is the first direct measurement
  of the size of a small main-belt binary.
  Although the interferometric observations of Isberga are at the edge
  of MIDI capabilities, the method described here is applicable to
  others suites of instruments (\eg, LBT, ALMA).
\end{abstract}

\begin{keyword}
%% keywords here, in the form: keyword \sep keyword
Asteroids, dynamics \sep Satellites of asteroids \sep  \sep Orbit determination
%% MSC codes here, in the form: \MSC code \sep code
%% or \MSC[2008] code \sep code (2000 is the default)

\end{keyword}

\end{frontmatter}

% \linenumbers

%% main text

%%%%%%%%%%%%%%%%%%%%%%%%%%%%%%%%%%%%%%%%%%%%%%%%%%%%%%%%%%%%%%%%%%%%%%%%%%%%%
%%% TAG %%%------%%%%%%%%%%%%%%%%%%%%%%%%%%%%%%%%%%%%%%%%%%%%%%%%%%%%%%%%%%%%
%%%%%%%%%%%%%%%%%%%%%%%%%%%%%%%%%%%%%%%%%%%%%%%%%%%%%%%%%%%%%%%%%%%%%%%%%%%%%
\section{Introduction}
  \indent Of the many properties that describe an asteroid, there is
  perhaps no quantity more fundamental to understand its composition and
  internal structure than its density. 
  With the exception of the fine-grained dust returned from asteroid (25\,143)
  Itokawa by the Hayabusa spacecraft
  \citep{2011-Science-333-Nakamura},
  \ben{our knowledge on the mineralogy of asteroids has been derived from
  remote-sensing photometry and spectroscopy in the visible and
  near-infrared, radar polarimetry, and comparison with meteorites studied in the
  laboratory \citep[e.g.,][]{2008-Icarus-195-Shepard, 2010-Icarus-207-Vernazza}.}
  These observables, however, tell us
  about surface composition only, which may or may not be
  reflective of the bulk composition of the body.
  \ben{The bulk density of meteorites spans a wide range, from the
  low-density ($\rho$\,$\sim$\,$1.6$\,g.cm$^{-3}$) primitive CI carbonaceous chondrite to the 
  dense       ($\rho$\,$\sim$\,$7.4$\,g.cm$^{-3}$) Hexahedrite iron meteorites
  \citep[see, \eg,][for meteorites density
    measurements]{1998-MPS-33-Consolmagno, 2008-ChEG-68-Consolmagno}.
  Comparison of asteroid bulk density with meteorite grain density
  provides a crude, yet useful, tool in the investigation of their bulk
  composition. This is particularly valuable for taxonomic types
  devoid of characteristic absorption bands in their spectrum, for
  which the analog meteorites cannot be ascertained otherwise.} \\
  \indent For asteroids with known surface mineralogy \ben{and 
    analog meteorite,} the density even allows
  us to make inference on the internal structure of the body.
  By comparing the \ben{grain}
  density of the surface material to the bulk density of the asteroid, we can
  detect the presence of denser material below the crust, like in the case of
  (4) Vesta \citep{2012-Science-336-Russell}, or
  the presence of \ben{large} voids, called macroprorosity,
  as for the rubble-pile (25\,143) Itokawa
  \citep{2006-Science-312-Fujiwara}.
  A recent comprehensive analysis of volume and mass determinations of
  about 300 asteroids has revealed clear differences of density and
  macroporosity among taxonomic types, 
  together with different trends with size and orbital populations
  \citep{2012-PSS-73-Carry}.
  This sample is, however, still limited in number and the precision
  of the majority of these estimates remains cruder than
  \ben{50\% (1-$\sigma$ cutoff)}. \\
  \indent In our quest for asteroid masses, the study of binary systems has
  been the most productive method 
  \citep{2012-PSS-73-Carry}.
  \ben{Spacecraft encounters provide the most precise mass determination
  \citep[at the percent level, \eg,][]{2011-Science-334-Paetzold}, but
  they will always remain limited to a few objects, while studies of orbit
  deflections during planetary encounters provide numerous mass
  estimates with limited precision 
  \citep[often above 50\% relative accuracy, see][for
    instance]{2011-AJ-142-Zielenbach, 2013-Icarus-222-Kuchynka}.} 
%  If spacecraft encounters provide the most
%  precise mass determination
%  \citep[\eg,][]{2011-Science-334-Paetzold}, they will always be limited to a very few
%  objects. At the other end of the spectrum, studies of orbit deflections during
%  asteroid close encounters and planetary ephemeris provide hundreds of mass
%  estimates, \ben{with mild precision only to date
%  \citep[often above 50\% relative accuracy, see][for instance]{2011-AJ-142-Zielenbach, 2013-Icarus-222-Kuchynka}}.
%
  With more than 200 binary systems known, and more discoveries announced almost
  monthly, the study of mutual orbits can provide numerous mass
  determinations. 
  For large separation binaries, where the companion can be imaged and tracked
  along its orbit \citep[\eg,][among others]{1999-Nature-401-Merline, 
    2005-Nature-436-Marchis, 2011-Icarus-211-Descamps,
    2011-AA-534-Carry, 2012-AA-543-Vachier}, the mass can be 
  determined to a high precision, \ben{typically about 10-15\%
    \citep{2012-Icarus-217-Carry}}.
%  However, the precision on the density remains  generally limited by the volume accuracy. 
  For the small binaries, detected and studied by the signature of mutual
  eclipses and occultations in their lightcurves,
  \ben{the density can be indirectly determined without measuring  
  the absolute size nor mass of the
  objects
  \citep[\eg,][]{2006-Icarus-181-Pravec,2012-Icarus-218-Pravec}. 
  This, however, requires to assume the same bulk density for both
  components \citep[\eg,][]{2009-Icarus-200-Scheirich}, which may be
  problematic if these small-sized binaries are formed by rotational
  breakup \citep{2008-Nature-454-Walsh}. The accuracy reached with
  this method can range from a few percent to 100\% depending on each
  system \citep{2012-Icarus-217-Carry}.} \\ 
  \indent We present here a suite of observations of the small
  \ben{main-belt} binary asteroid (939) Isberga
  (orbital elements: a=2.246 au, e=0.177, i=2.588\degr)
  aiming at determining its surface composition, mutual orbit,
  mass, diameter, density, and macroporosity.
  We describe in Section~\ref{sec: obs} the different methods of observation
  we use, we present the analysis of the surface composition of
  Isberga in Section~\ref{sec: surface} and of the physical properties
  of the system in Section~\ref{sec: orbit}.

%%%%%%%%%%%%%%%%%%%%%%%%%%%%%%%%%%%%%%%%%%%%%%%%%%%%%%%%%%%%%%%%%%%%%%%%%%%%%
%%% TAG %%%------%%%%%%%%%%%%%%%%%%%%%%%%%%%%%%%%%%%%%%%%%%%%%%%%%%%%%%%%%%%%
%%%%%%%%%%%%%%%%%%%%%%%%%%%%%%%%%%%%%%%%%%%%%%%%%%%%%%%%%%%%%%%%%%%%%%%%%%%%%
\section{Observations and data reduction\label{sec: obs}}
%
%  \indent We describe below the different observing campaigns we set up to
%  study the orbital and surface properties of (939) Isberga.

  \subsection{Optical lightcurves\label{sec: LC}}
\begin{figure}[t]
\begin{center}
  \includegraphics[width=.49\textwidth]{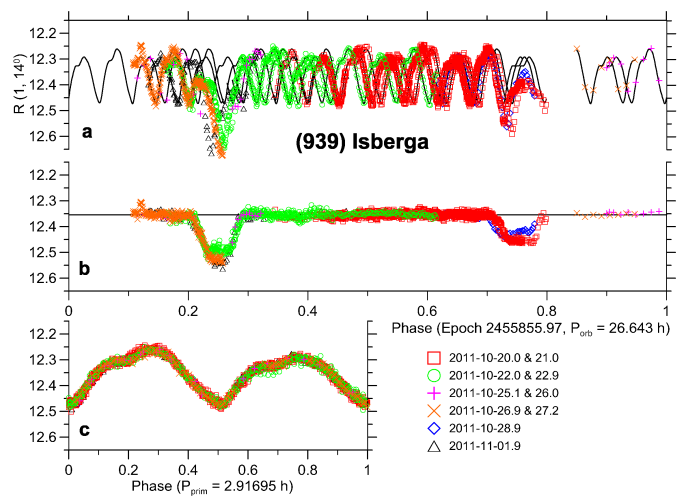}%
  \caption[Orbital and rotation lightcurve of Isberga]{%
  \label{fig: lc1}
  Lightcurves of Isberga showing the mutual eclipses and photometric
  variability induced by the primary rotation.
  \textbf{a:} All the lightcurves acquired between
  \ben{2011, 
  October the \nth{20} and 2011, November the \nth{1}} folded over the
  synodic orbital period of 26.643\,h. 
  \textbf{b:} The same as above, with the orbital component of the lightcurve only.
  \textbf{c:} The rotation component of the lightcurve only, folded over the
  rotation period of 2.91695\,h.
  }
\end{center}
\end{figure}

    \indent The binarity of Isberga was reported by 
    \citet{2008-MPBu-35-Molnar} from
    optical lightcurves obtained over 6 nights in 2006 at
    the Calvin-Rehoboth Observatory. 
    \ben{The rotation period of the primary 
      and the orbital period for the satellite were determined 
      to 2.9173\,$\pm$\,0.0003\,h
      and 26.8\,$\pm$\,0.1\,h.}
    We report observations carried out during 
    2 nights from the 2008/2009 opposition, \ben{43 nights in 2010,
      54 nights in 2011, and 2 nights in 2012}. 
    \ben{We provide a detailed list of all the lightcurves with ancillary 
    information in Table~\ref{tab: lc}. A subset of the lightcurves is
    plotted in Fig.~\ref{fig: lc1}, showing evidences for mutual
    events.} \\
%
%
%    \indent The apparition of 2011 was also followed by many different observers.
%    We observed Isberga with the Calar Alto 1.2\,m telescope over 18 nights in
%    August, September, October, and December 2011. The call of observations we released
%    at that time was largely followed by both professional and amateur
%    astronomers, \eg,
%    12 nights at the
%    Astronomical Observatory of the Autonomous Region of the Aosta Valley (OAVdA), 
%    2 nights at the Borowiec observatory, 
%    11 nights with Trappist at La Silla Observatory, 
%    5 nights at the Michel Bonnardeau's observatory (MBCAA),
%    and 18 nights from the ``Courbe de Rotation'' (CdR) amateur group from different
%    locations. 
%    We provide a detailed list of all the lightcurves with ancillary 
%    information in Table~\ref{tab: lc}. 
%    As additional lightcurves were gathered, the 
%    ephemeris of the mutual events were updated, leading to many
%    successful observations of the mutual eclipses between Isberga and its
%    companion over the fall 2011 (Fig.~\ref{fig: lc1}). \\
%
    \indent As many observers acquired lightcurves of Isberga, we do not
    go here into the specifics of the data reduction and photometry
    measurements used by each.
    Standard procedures were used to reduce the data,
    including bad pixel removal, bias subtraction, and flat-field correction.
    Aperture photometry was used to measure the relative flux of Isberga with
    surrounding stars and build its lightcurves. 
    \ben{In lightcurve decomposition, the magnitude scale zero points of
    individual nights (sessions) were taken as free parameters. Their
    uncertainties were generally less than 0.01 mag and we checked by
    experimenting with them that they did not add a significant
    uncertainty in subsequent modeling of the system, and we did not
    propagate them there.}

%%%%%%------ FIGURE --- Begin --- LC1 ------%%%%%%
%%%%%%------ FIGURE ---  End  --- LC1 ------%%%%%%
%
%
%%%%%%%------ TABLE --- Begin --- LC Observations ------%%%%%%
\begin{table*}[t]
\begin{center}
%%%---Caption and Labeling ---%%%
  \caption[Lightcurve observations]{
    \label{tab: lc}
    Date, duration ($\mathcal{D}$), phase angle ($\alpha$), observatory, and observers
    of each lightcurve used in present study. The observatory code are IAU
    codes \ben{
    (G98: Calvin-Rehoboth Observatory,
     493: Calar Alto 1.2\,m telescope, 
     I40: TRAPPIST at La Silla Observatory,
     J23: Centre Astronomique de La Couy{\`e}re, 
     187: Borowiec observatory, 
     634: Crolles observatory,
     586: T60 and T1M at Pic du Midi),
     except for Far, MBo, StB, and VFa that correspond to 
    the Farigourette, Michel Bonnardeau's (MBCAA), Saint Barthelemy,
    and Villefagnan observatories. }
}
%%%--- Table Content ---%%%
  \begin{tabular}{ccrcl|ccrcl}
    \hline\hline
    Date & $\mathcal{D}$ & $\alpha$ & Obs. & Observers &
    Date & $\mathcal{D}$ & $\alpha$ & Obs. & Observers \\
    (UT) & (h) & (\degr) & & &
    (UT) & (h) & (\degr) & & \\
    \hline
2006-02-24 -- 09:36 & 5.3 & 11.7 & G98 & \footnotesize{Molnar et al.       } & 2011-09-06 -- 14:24 & 6.0 & 23.8 & 493 & \footnotesize{Mottola, Hellmich   } \\
2006-02-26 -- 14:24 & 6.8 & 12.5 & G98 & \footnotesize{Molnar et al.       } & 2011-09-07 -- 16:48 & 6.3 & 23.3 & 493 & \footnotesize{Mottola, Hellmich   } \\
2006-02-27 -- 16:48 & 6.0 & 12.9 & G98 & \footnotesize{Molnar et al.       } & 2011-09-19 -- 21:36 & 0.4 & 18.0 & I40 & \footnotesize{Jehin et al.        } \\
2006-02-28 -- 19:12 & 7.0 & 13.3 & G98 & \footnotesize{Molnar et al.       } & 2011-09-25 -- 12:00 & 6.4 & 15.1 & G98 & \footnotesize{Molnar et al.       } \\
2006-03-03 -- 07:12 & 4.5 & 14.2 & G98 & \footnotesize{Molnar et al.       } & 2011-09-26 -- 14:24 & 7.5 & 14.5 & G98 & \footnotesize{Molnar et al.       } \\
2006-03-04 -- 09:36 & 5.1 & 14.6 & G98 & \footnotesize{Molnar et al.       } & 2011-09-27 -- 16:48 & 6.2 & 13.9 & G98 & \footnotesize{Molnar et al.       } \\
2008-12-31 -- 02:24 & 1.3 &  6.2 & G98 & \footnotesize{Molnar et al.       } & 2011-09-28 -- 19:12 & 1.9 & 13.3 & G98 & \footnotesize{Molnar et al.       } \\
2009-01-01 -- 02:24 & 1.3 &  6.7 & G98 & \footnotesize{Molnar et al.       } & 2011-10-02 -- 04:48 & 1.5 & 11.3 & I40 & \footnotesize{Jehin et al.        } \\
2010-03-25 -- 12:00 & 5.2 &  3.6 & G98 & \footnotesize{Molnar et al.       } & 2011-10-07 -- 16:48 & 2.4 &  8.1 & I40 & \footnotesize{Jehin et al.        } \\
2010-03-28 -- 19:12 & 6.6 &  2.1 & G98 & \footnotesize{Molnar et al.       } & 2011-10-08 -- 19:12 & 3.7 &  7.4 & I40 & \footnotesize{Jehin et al.        } \\
2010-04-03 -- 07:12 & 6.4 &  1.5 & G98 & \footnotesize{Molnar et al.       } & 2011-10-09 -- 21:36 & 4.2 &  6.8 & I40 & \footnotesize{Jehin et al.        } \\
2010-04-07 -- 16:48 & 6.6 &  3.3 & G98 & \footnotesize{Molnar et al.       } & 2011-10-19 -- 21:36 & 4.7 &  2.3 & J23 & \footnotesize{Montier             } \\
2010-04-08 -- 19:12 & 6.9 &  3.8 & G98 & \footnotesize{Molnar et al.       } & 2011-10-20 -- 00:00 & 8.5 &  2.3 & VFa & \footnotesize{Barbotin, Behrend   } \\
2010-04-09 -- 21:36 & 7.0 &  4.4 & G98 & \footnotesize{Molnar et al.       } & 2011-10-20 -- 00:00 & 8.3 &  2.3 & VFa & \footnotesize{Barbotin, Behrend   } \\
2010-04-12 -- 04:48 & 7.0 &  5.4 & G98 & \footnotesize{Molnar et al.       } & 2011-10-21 -- 02:24 & 7.5 &  2.4 & Far & \footnotesize{Morelle, Behrend    } \\
2010-04-13 -- 07:12 & 5.7 &  6.0 & G98 & \footnotesize{Molnar et al.       } & 2011-10-21 -- 02:24 & 2.9 &  2.4 & I40 & \footnotesize{Jehin et al.        } \\
2010-04-14 -- 09:36 & 6.0 &  6.5 & G98 & \footnotesize{Molnar et al.       } & 2011-10-22 -- 04:48 & 9.3 &  2.6 & VFa & \footnotesize{Barbotin, Behrend   } \\
2010-04-16 -- 14:24 & 4.6 &  7.5 & G98 & \footnotesize{Molnar et al.       } & 2011-10-22 -- 04:48 & 1.5 &  2.6 & J23 & \footnotesize{Montier             } \\
2010-04-18 -- 19:12 & 5.9 &  8.5 & G98 & \footnotesize{Molnar et al.       } & 2011-10-22 -- 04:48 & 5.8 &  2.6 & 187 & \footnotesize{Marciniak et al.    } \\
2010-04-18 -- 19:12 & 2.4 &  8.5 & G98 & \footnotesize{Molnar et al.       } & 2011-10-22 -- 04:48 & 7.4 &  2.6 & MBo & \footnotesize{Bonnardeau          } \\
2010-04-23 -- 07:12 & 5.0 & 10.5 & G98 & \footnotesize{Molnar et al.       } & 2011-10-22 -- 04:48 & 5.0 &  2.6 & 634 & \footnotesize{Farissier           } \\
2010-04-23 -- 07:12 & 3.8 & 10.5 & G98 & \footnotesize{Molnar et al.       } & 2011-10-25 -- 12:00 & 1.3 &  4.1 & 493 & \footnotesize{Mottola, Hellmich   } \\
2010-05-04 -- 09:36 & 5.0 & 15.0 & G98 & \footnotesize{Molnar et al.       } & 2011-10-26 -- 14:24 & 8.5 &  4.7 & 493 & \footnotesize{Mottola, Hellmich   } \\
2010-05-05 -- 12:00 & 4.8 & 15.4 & G98 & \footnotesize{Molnar et al.       } & 2011-10-26 -- 14:24 & 2.6 &  4.7 & 493 & \footnotesize{Mottola, Hellmich   } \\
2010-05-07 -- 16:48 & 2.2 & 16.2 & G98 & \footnotesize{Molnar et al.       } & 2011-10-27 -- 16:48 & 4.0 &  5.3 & I40 & \footnotesize{Jehin et al.        } \\
2010-05-08 -- 19:12 & 4.5 & 16.6 & G98 & \footnotesize{Molnar et al.       } & 2011-10-28 -- 19:12 & 4.4 &  5.9 & StB & \footnotesize{Carbognani          } \\
2010-05-09 -- 21:36 & 4.8 & 17.0 & G98 & \footnotesize{Molnar et al.       } & 2011-11-01 -- 02:24 & 4.9 &  7.7 & StB & \footnotesize{Carbognani          } \\
2010-05-10 -- 00:00 & 4.3 & 17.0 & G98 & \footnotesize{Molnar et al.       } & 2011-11-01 -- 02:24 & 8.0 &  7.7 & 493 & \footnotesize{Mottola, Hellmich   } \\
2010-05-11 -- 02:24 & 0.5 & 17.3 & G98 & \footnotesize{Molnar et al.       } & 2011-11-15 -- 12:00 & 5.7 & 15.1 & StB & \footnotesize{Carbognani          } \\
2010-05-15 -- 12:00 & 3.1 & 18.7 & G98 & \footnotesize{Molnar et al.       } & 2011-11-15 -- 12:00 & 3.9 & 15.1 & Far & \footnotesize{Morelle, Behrend    } \\
2010-05-16 -- 14:24 & 0.5 & 19.0 & G98 & \footnotesize{Molnar et al.       } & 2011-11-16 -- 14:24 & 5.8 & 15.6 & StB & \footnotesize{Carbognani          } \\
2010-05-17 -- 16:48 & 2.1 & 19.3 & G98 & \footnotesize{Molnar et al.       } & 2011-11-16 -- 14:24 & 7.1 & 15.6 & Far & \footnotesize{Morelle, Behrend    } \\
2010-05-19 -- 21:36 & 3.7 & 19.9 & G98 & \footnotesize{Molnar et al.       } & 2011-11-17 -- 16:48 & 8.0 & 16.1 & Far & \footnotesize{Morelle, Behrend    } \\
2010-05-20 -- 00:00 & 2.7 & 19.9 & G98 & \footnotesize{Molnar et al.       } & 2011-11-18 -- 19:12 & 6.1 & 16.6 & StB & \footnotesize{Carbognani          } \\
2010-05-21 -- 02:24 & 0.4 & 20.2 & G98 & \footnotesize{Molnar et al.       } & 2011-11-20 -- 00:00 & 5.8 & 17.1 & StB & \footnotesize{Carbognani          } \\
2010-05-28 -- 19:12 & 2.7 & 21.9 & G98 & \footnotesize{Molnar et al.       } & 2011-11-22 -- 04:48 & 6.6 & 18.0 & StB & \footnotesize{Carbognani          } \\
2010-05-29 -- 21:36 & 3.2 & 22.1 & G98 & \footnotesize{Molnar et al.       } & 2011-11-26 -- 14:24 & 8.7 & 19.6 & 586 & \footnotesize{Fauvaud et al.      } \\
2010-05-30 -- 00:00 & 1.4 & 22.1 & G98 & \footnotesize{Molnar et al.       } & 2011-11-27 -- 16:48 & 5.3 & 20.0 & 586 & \footnotesize{Fauvaud et al.      } \\
2010-05-31 -- 02:24 & 1.8 & 22.4 & G98 & \footnotesize{Molnar et al.       } & 2011-12-01 -- 02:24 & 0.9 & 21.1 & I40 & \footnotesize{Jehin et al.        } \\
2010-06-01 -- 02:24 & 2.6 & 22.5 & G98 & \footnotesize{Molnar et al.       } & 2011-12-03 -- 07:12 & 2.4 & 21.8 & I40 & \footnotesize{Jehin et al.        } \\
2010-06-02 -- 04:48 & 2.3 & 22.7 & G98 & \footnotesize{Molnar et al.       } & 2011-12-04 -- 09:36 & 1.2 & 22.1 & I40 & \footnotesize{Jehin et al.        } \\
2010-06-03 -- 07:12 & 2.6 & 22.9 & G98 & \footnotesize{Molnar et al.       } & 2011-12-11 -- 02:24 & 2.0 & 23.8 & I40 & \footnotesize{Jehin et al.        } \\
2010-06-04 -- 09:36 & 2.4 & 23.1 & G98 & \footnotesize{Molnar et al.       } & 2011-12-18 -- 19:12 & 5.0 & 25.3 & 493 & \footnotesize{Mottola, Hellmich   } \\
2010-06-08 -- 19:12 & 2.3 & 23.7 & G98 & \footnotesize{Molnar et al.       } & 2011-12-19 -- 21:36 & 5.7 & 25.4 & 493 & \footnotesize{Mottola, Hellmich   } \\
2011-08-29 -- 21:36 & 4.6 & 26.4 & 493 & \footnotesize{Mottola, Hellmich   } & 2011-12-20 -- 00:00 & 6.1 & 25.5 & 493 & \footnotesize{Mottola, Hellmich   } \\
2011-08-30 -- 00:00 & 5.0 & 26.4 & 493 & \footnotesize{Mottola, Hellmich   } & 2011-12-21 -- 02:24 & 5.7 & 25.6 & 493 & \footnotesize{Mottola, Hellmich   } \\
2011-08-31 -- 02:24 & 5.7 & 26.0 & 493 & \footnotesize{Mottola, Hellmich   } & 2011-12-22 -- 04:48 & 6.0 & 25.8 & 493 & \footnotesize{Mottola, Hellmich   } \\
2011-09-03 -- 07:12 & 4.6 & 25.0 & 493 & \footnotesize{Mottola, Hellmich   } & 2011-12-23 -- 07:12 & 5.1 & 25.9 & 493 & \footnotesize{Mottola, Hellmich   } \\
2011-09-04 -- 09:36 & 3.1 & 24.6 & 493 & \footnotesize{Mottola, Hellmich   } & 2012-01-18 -- 19:12 & 4.6 & 27.4 & 586 & \footnotesize{Vachier, Colas, Lecacheux} \\
2011-09-05 -- 12:00 & 6.2 & 24.2 & 493 & \footnotesize{Mottola, Hellmich   } & 2012-01-20 -- 00:00 & 4.0 & 27.3 & 586 & \footnotesize{Vachier, Colas, Lecacheux} \\
    \hline
  \end{tabular}
\end{center}
\end{table*}
%%%%%%%------ TABLE ---  End  --- LC Observations ------%%%%%%

  \subsection{Near-infrared spectroscopy\label{sec: spectro}}
\begin{figure}[t]
\begin{center}
  \includegraphics[width=.49\textwidth]{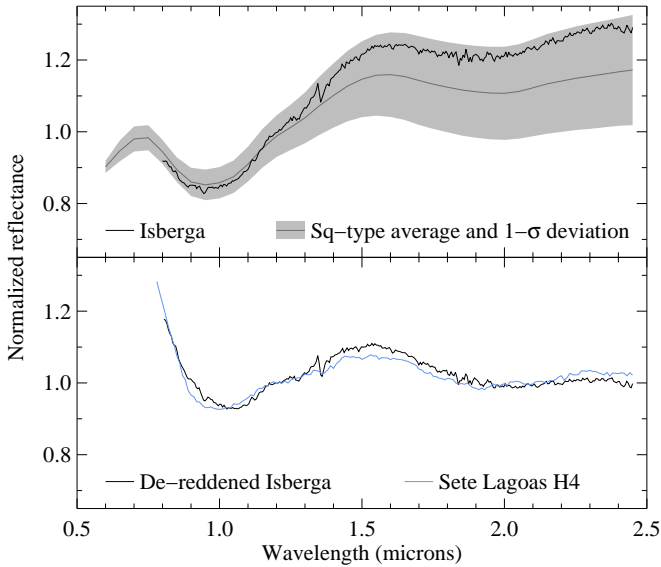}%
  \caption[Near-infrared spectrum of Isberga]{%
  \label{fig: spec}
  \ben{
  \textbf{Top:}
  Near-infrared spectrum of Isberga normalized at 1.20 $\mu$m compared
  with the average Bus-DeMeo Sq-type spectrum.
  \textbf{Bottom:} De-reddened spectrum of Isberga
  \ben{\citep[using the space weathering model of][see Sec.~\ref{sec:
      surface}]{2006-Icarus-184-Brunetto}} compared 
  with the ordinary chondrite H4 Sete Lagoas (RELAB sample ID: MH-JFB-021).
  }
  }
\end{center}
\end{figure}

    \indent To constrain the surface mineralogy, 
    we acquired a near-infrared spectrum of Isberga 
    on \ben{2011, August the \nth{22}, at a phase angle of 28\degr,}
    as part of
    the MIT-Hawaii-IRTF joint campaign for NEO reconnaissance
    \citep{2006-LPI-37-Binzel}.
    Data from this survey are publicly
    available at \texttt{smass.mit.edu}.
    Observations were taken on the 3-meter NASA
    Infrared Telescope Facility at the Mauna Kea Observatory. We used the
    instrument SpeX \citep{2003-PASP-115-Rayner},
    a near-infrared spectrograph in low resolution mode over 0.8 to
    2.5\,$\mu$m. \\
    \indent Isberga was observed near the meridian \ben{(airmass\,$<$\,1.3)} in two
    different positions, here denoted A and B, on a 0.8\,$\times$\,15 arcsecond$^2$
    slit aligned north-south. Exposure times were 120 seconds,
    and we measured 4 A-B pairs. Solar analog stars were
    observed at similar airmass throughout the night to correct for telluric
    absorption. We used the same set of
    solar analogs as the SMASS program
    \citep{2004-Icarus-170-Binzel, 2006-LPI-37-Binzel} that have been
    in use for over a decade. \\
    \indent Data reduction and spectral extraction were performed using the
    Image Reduction and Analysis Facility \citep[IRAF,][]{1993-ASPC-52-Tody}
    provided by the National Optical Astronomy Observatories (NOAO).
    Correction in regions with strong
    telluric absorption was performed in IDL using an atmospheric transmission
    (ATRAN) model by \citet{1992-NASA-Lord}. 
    More detailed information on the observing and reduction procedures
    can be found in
    \citet{2004-Icarus-172-Rivkin} and
    \citet{2008-Icarus-194-DeMeo}.
    We present the resulting spectrum of Isberga in Fig.~\ref{fig: spec}.

%%%%%%------ FIGURE --- Begin --- Spectrum ------%%%%%%
%%%%%%------ FIGURE ---  End  --- Spectrum ------%%%%%%

  \subsection{Mid-infrared interferometry\label{sec: vlti}}
\begin{figure}[t]
\begin{center}
  \includegraphics[width=.49\textwidth]{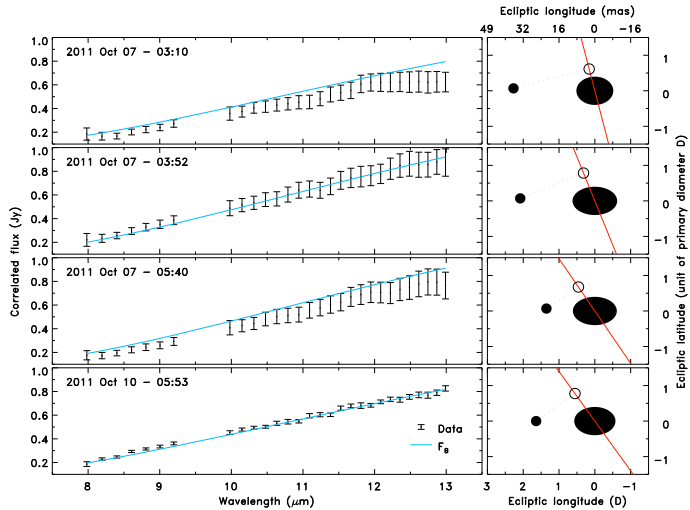}%
  \caption[Mid-infrared correlated flux of Isberga]{%
  \label{fig: midi}
  \textbf{Left:}
  Correlated flux of Isberga observed with MIDI over the four epochs listed in
  Table~\ref{tab: midi}.
  The best-fit solution of binary model ($F_B$) is also plotted as
  a solid \ben{blue} line. 
  \textbf{Right:}
  Corresponding geometry of the system on the plane of the sky.
  The red line represents the \ben{projected VLTI baseline},
  the black ellipse Isberga, the black disk its satellite, and the
  black circle the projection of the satellite on the baseline.
  }
\end{center}
\end{figure}

 \indent \ben{Mid-infrared interferometry can provide direct
   measurements of the angular extension of asteroids 
   \citep{2009-ApJ-694-Delbo, 2011-Icarus-215-Matter,
     2013-Icarus-226-Matter}.
   We used the MID-infrared Interferometric instrument (MIDI) of the Very
   Large Telescope Interferometer (VLTI),
   combining two of the 8.2\,m Unitary Telescopes, UT1 and UT2, with a
   baseline separation of 57\,m, providing a 
   high angular resolution of 
   $\frac{\lambda}{2B}$\,$\approx$\,0.02\arcsec~at $\lambda \approx
   10$\,$\mu$m, corresponding to about 10 km projected
   at the distance of Isberga at the time of observation.}\\
   \indent \ben{MIDI records the interference fringes between
   two beams of incoming light, which gives access to the complex
   degree of coherence (or complex visibility) between the
   beams. This complex visibility is the Fourier transform of the object
   brightness distribution on the plane of the sky, stacked along 
   the baseline direction and sampled at the spatial frequency
   $\boldsymbol{\boldsymbol{B}/\lambda}$, where
   $\boldsymbol{\boldsymbol{B}}$ is the baseline vector. In this
   work, we focused on the correlated flux observable, which is the
   modulus of the complex visibility.} \\ 
   \indent Fringes on Isberga were acquired at three observing epochs
    in visitor mode on \ben{2011, October the \nth{6}}, and at four observing epochs in service mode
    on \ben{2011, October the \nth{10}} (Table~\ref{tab: midi}), \ben{following the
      observing procedure of \citet{2004-AA-423-Leinert}}. 
    The fringes were dispersed using the prism of MIDI, which
    provides a spectral resolving power of
    $\lambda/\Delta \lambda$\,$\approx$\,30
    at $\lambda$\,$=$\,10\,$\mu$m.
    \ben{One correlated flux measurement, dispersed
    over the N-band (from 8 to 13~$\mu$m), was obtained
    for each observing epoch.}  
    Our observations also included a photometric and 
    interferometric calibrator star, HD~15396, \ben{to determine the
      atmospheric and instrumental transfer function. Our calibrator
      was chosen to be angularly unresolved,} and to have a minimum
    angular separation with the source ($\approx$3\degr) and 
    a similar airmass (see Table~\ref{tab: midi}). \\
    \indent \ben{The correlated flux measurements of (939)
    Isberga were extracted using the data reduction software package EWS \citep[Expert
    WorkStation, see][for a detailed description]{2004-SPIE-Jaffe}.}
    Using the closest calibrator
    observation in time, calibrated correlated fluxes of (939) Isberga were
    obtained by multiplying the ratio target/calibration star raw correlated flux
    counts by the absolutely calibrated infrared flux of the
    calibrator \ben{\citep[see][for a complete
    description of the data reduction and calibration
      procedure]{2011-Icarus-215-Matter, 2013-Icarus-226-Matter}}.\\
    \indent \ben{Uncertainties on the correlated flux are estimated
      considering two contributions.
      First, a short timescale effect
      (much shorter than typical observations of $\approx$2\,min), 
      dominated by photon noise from the object and thermal background
      emission. This statistical uncertainty is estimated by splitting a
      complete exposure, consisting of several thousand of frames and
      leading to one correlated flux measurement, into five equal
      parts and by deriving their standard deviation for every
      spectral channel.
      Second, the slow variations in the flux
      transmission of the atmosphere and/or variations of the thermal
      background can introduce offsets between repeated observations
      across the night.% (negligible for a single observation or exposure).
      A rough estimate of this offset-like contribution is
      obtained by calibrating each correlated flux 
      measurement against all the calibrators of the night, and then computing the standard deviation
      \citep[see][]{2007-NewAR-51-Chesneau}. Such estimate was
      only possible for the three measurements of 2011, October the \nth{6}
      where the error bars correspond to the quadratic sum of these
      two sources of uncertainty.}\\ % (see Fig.~\ref{fig: midi}).}\\ 
    \indent \ben{The four fringe measurements on 2011, October the
      \nth{10} were acquired over a 
      period of 15 minutes with only one calibrator observation. Since
      this is short compared to the estimated 
      rotation and orbital period of Isberga of 2.9\,h and 26.8\,h respectively,
      the system apparent geometry, which dominates the data compared
      to, e.g., putative surface composition heterogeneity, did not change.
      We thus averaged the four observing epochs to reduce the 
      statistical noise. Assuming that the averaging process also
      removed the possible offsets affecting the four measurements, the
      corresponding error bars only include the ``averaged'' short-term
      statistical error contribution. 
      Fig.~\ref{fig: midi} shows the
      four measurements resulting from the seven initial individual MIDI
      measurements listed in Table~\ref{tab: midi}.}

%
%%%%%%%------ TABLE --- Begin --- MIDI Observations ------%%%%%%
\begin{table*}[t]
\begin{center}
%%%---Caption and Labeling ---%%%
  \caption[Mid-infrared interferometric observations]{
    \label{tab: midi}
    projected baseline 
    (length $B$ and position angle $PA$ counted from North to East),
    seeing, and airmass 
    for each observation of Isberga and its calibrator (labeled in the last
    column) using MIDI on the UT1-UT2 baseline of the VLTI.
}
%%%--- Table Content ---%%%
 \begin{tabular}[!ht]{lcccccc}
 \hline
 \multicolumn{1}{c}{Object} & Date & $B$ & $PA$ & Seeing & Airmass & Label\\
 & (UT) & (m) & (\degr) & (\arcsec) \\
  \hline
  {\small (939) Isberga} & 2011-10-07 03:09:43 & 37.4 & 14.0 & 0.53 & 1.60 & 1\\
  {\small HD~13596}      & 2011-10-07 03:34:14 & 37.2 & 13.0 & 0.70 & 1.63 &  {\small Calib}\\
  {\small (939) Isberga} & 2011-10-07 03:51:45 & 39.7 & 21.6 & 0.76 & 1.43 & 2\\
  {\small HD~13596}      & 2011-10-07 04:10:16 & 39.0 & 19.8 & 1.09 & 1.47 & {\small Calib}\\
  {\small (939) Isberga} & 2011-10-07 05:40:04 & 47.5 & 33.7 & 0.72 & 1.30 & 3\\
  {\small HD~13596}      & 2011-10-07 06:00:29 & 46.8 & 33.1 & 0.84 & 1.30 & {\small Calib}\\
  {\small (939) Isberga} & 2011-10-10 05:44:02 & 48.9 & 34.7 & 0.70 & 1.31 & 4\\
  {\small (939) Isberga} & 2011-10-10 05:54:27 & 49.6 & 35.2 & 0.70 & 1.33 & 4\\
  {\small (939) Isberga} & 2011-10-10 05:57:56 & 49.8 & 35.3 & 0.70 & 1.33 & 4\\
  {\small (939) Isberga} & 2011-10-10 06:01:53 & 50.0 & 35.5 & 0.76 & 1.34 & 4\\
  {\small HD~13596}      & 2011-10-10 06:19:44 & 49.0 & 35.0 & 0.79 & 1.32 & {\small Calib}\\
  \hline
 \end{tabular}
\end{center}
\end{table*}
%%%%%%%------ TABLE ---  End  --- MIDI Observations ------%%%%%%

%%%%%%------ FIGURE --- Begin --- MIDI ------%%%%%%
%%%%%%------ FIGURE ---  End  --- MIDI ------%%%%%%

%%%%%%%%%%%%%%%%%%%%%%%%%%%%%%%%%%%%%%%%%%%%%%%%%%%%%%%%%%%%%%%%%%%%%%%%%%%%%
%%% TAG %%%------%%%%%%%%%%%%%%%%%%%%%%%%%%%%%%%%%%%%%%%%%%%%%%%%%%%%%%%%%%%%
%%%%%%%%%%%%%%%%%%%%%%%%%%%%%%%%%%%%%%%%%%%%%%%%%%%%%%%%%%%%%%%%%%%%%%%%%%%%%
\section{Surface composition\label{sec: surface}}
  \indent We use the Virtual Observatory (VO) service
  M4AST\footnote{\href{http://m4ast.imcce.fr/}{http://m4ast.imcce.fr/}}
  \citep{2012-AA-544-Popescu} to analyze the near-infrared
  reflectance spectrum of Isberga shown in Fig.~\ref{fig: spec}.
  It presents two broad absorption bands centered at
  \ben{0.95\,$\pm$\,0.01}
  and
  \ben{1.91\,$\pm$\,0.01}\,$\mu$m, tracers of olivine and pyroxene assemblages.
  We classify Isberga as an S- or Sq-type asteroid
  \citep[in the classification scheme by][]{2009-Icarus-202-DeMeo},
  the main class in the inner part
  of the asteroid belt \citep{2013-Icarus-226-DeMeo,
    2014-Nature-505-DeMeo}.
  We also use M4AST to determine the degree of space weathering
  presented by Isberga's surface\ben{, following the space weathering
    model by
  \citet{2006-Icarus-184-Brunetto}, valid for pyroxenes and olivines
  \citep[see, e.g.,][]{2007-AA-472-Brunetto, 2009-Nature-458-Vernazza,
    2012-AA-537-Fulvio}.
  In this model, the effect of space weathering is a multiplicative
  exponential continuum written as $K e^{(C_S / \lambda)}$.
  This model is based on many laboratory experiments on ordinary
  chondrites and mimic the effect of space weathering on lunar-like
  surfaces \citep[increased spectral slope and decreased albedo,
    see][among others]{2001-Nature-410-Sasaki, 2004-AREPS-32-Chapman, 
    2005-Icarus-174-Strazzulla}. Space weathering trends are more
  subtle and complex for asteroids within the S-complex, owing to the
  different compositions it encompass \citep{1993-Icarus-106-Gaffey,
    2010-Icarus-209-Gaffey}, albeit spectral reddening is consistent.}\\
  \indent \ben{We determine a reddening strength of $C_S$\,=\,-0.6\,$\mu$m, a value
    similar to, \eg, (158) Koronis 
    (measured with M4AST on the near-infrared spectrum by 
    \citet{2002-Icarus-159-Burbine} obtained at a phase angle of 19\degr) and
    corresponding to significant weathering (responsible for the
    higher spectral slope of Isberga compared with the average Sq-class of
    \citet{2009-Icarus-202-DeMeo} in Fig.~\ref{fig: spec}).
    The spectrum of Isberga was however obtained at a large phase
    angle of 28\degr~(Sec.~\ref{sec: spectro}), and part of the
    reddening may be caused by the observing geometry.
    Spectral observations of Isberga at visible wavelengths and small
    phase angle will help refine its taxonomic classification and
    state of space weathering. }\\ 
  \indent \ben{We determine a visible geometric albedo of
  $p_V$\,=\,0.14$^{+0.09}_{-0.06}$
  (Sec.~\ref{sec: orbmidi}) which is lower, yet consistent,} 
  than the average albedo of asteroids in the S-complex
  \ben{\citep[0.197\,$\pm$\,0.153, see][for values based on 
    WISE mid-infrared surveys]{2012-Icarus-221-Pravec} and corresponds to
    the first quartile of all 
  Bus-DeMeo S-complex asteroids \citep[based on Fig.~6
    by][]{2011-ApJ-741-Mainzer}.}
%  This value is consistent with the trend of weathered
%  S-types to be darker than ``fresh'' surfaces \citep[see, \eg, Fig.~1
%    in][]{2006-Icarus-184-Brunetto} and supports this classification. 
  \ben{We finally search for the best-fit (M4AST $\chi^2$ match) meteorite in the
  Relab spectral database to Isberga spectrum, corrected from the
  reddening (either due to the  phase angle or space weathering}. Ordinary
  chondrites provide the most-promising candidates, as to be expected
  from the Sq-type classification, and the 
  best-match is found for the H4 Sete Lagoas (sample MH-JFB-021).

%%%%%%%%%%%%%%%%%%%%%%%%%%%%%%%%%%%%%%%%%%%%%%%%%%%%%%%%%%%%%%%%%%%%%%%%%%%%%
%%% TAG %%%------%%%%%%%%%%%%%%%%%%%%%%%%%%%%%%%%%%%%%%%%%%%%%%%%%%%%%%%%%%%%
%%%%%%%%%%%%%%%%%%%%%%%%%%%%%%%%%%%%%%%%%%%%%%%%%%%%%%%%%%%%%%%%%%%%%%%%%%%%%
\section{Mutual orbit: size, mass, and density\label{sec: orbit}}
  \indent We describe here the different steps that lead to the determination
  of the geometric properties of the binary Isberga, \eg, component diameter ratio,
  semi-major axis of the orbit, absolute size.

  \subsection{Lightcurve analysis and orbit determination\label{sec: orblc}}
\begin{figure}[ht]
\begin{center}
  \includegraphics[width=.35\textwidth]{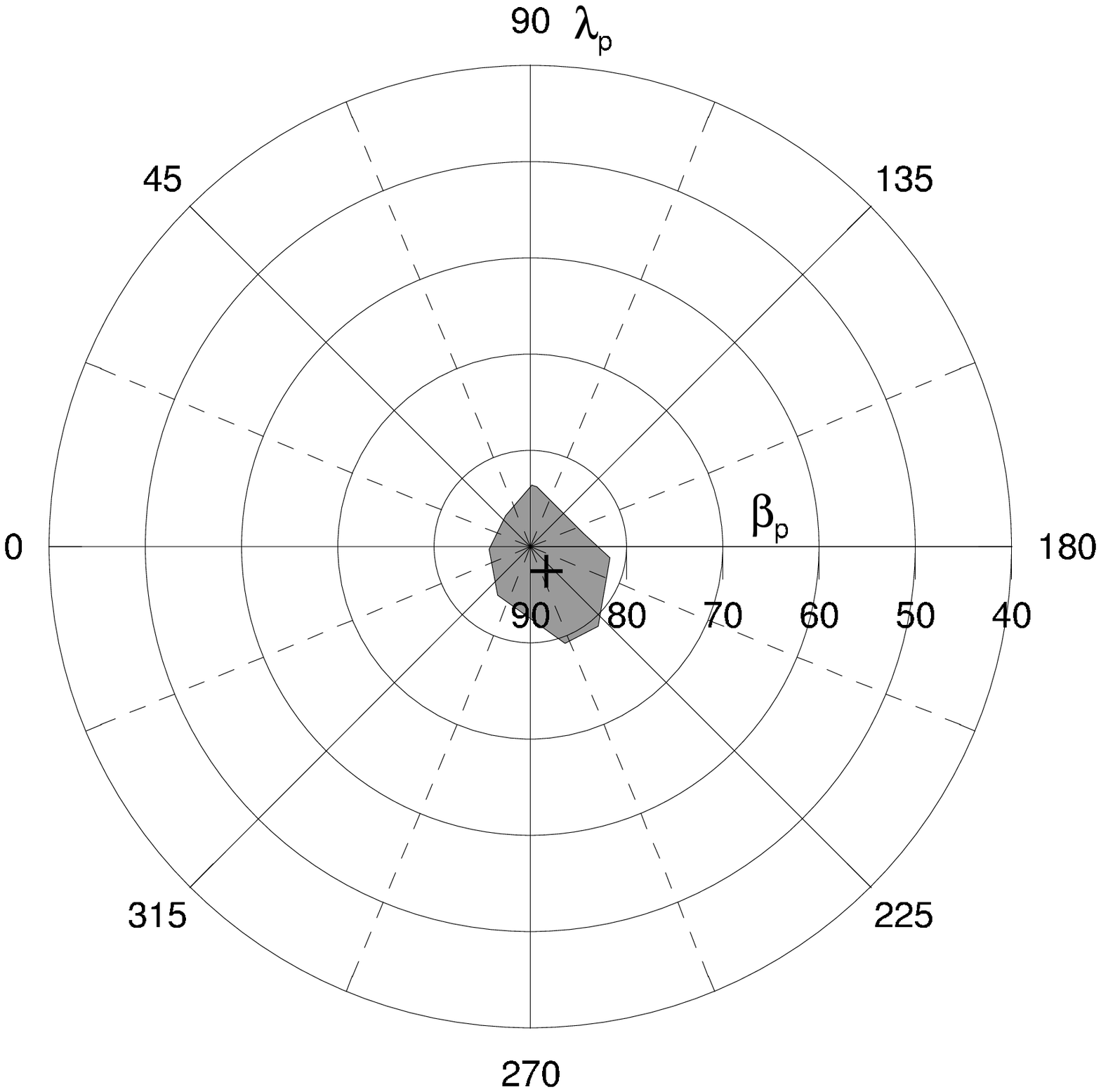}%
  \caption[Location of the spin of Isberga]{%
  \label{fig: spin}
  The 3-$\sigma$ confidence interval for the pole of the mutual orbit
  of Isberga (grey area) on an ECJ2000 grid, which can be \ben{approximated} by an
  ellipse of semi-major axes of 8\degr~and 6\degr, respectively. The north
  pole of the asteroid's heliocentric orbit is marked 
  with the black cross.
  }
\end{center}
\end{figure}

\begin{figure}[t]
\begin{center}
  \includegraphics[width=.49\textwidth]{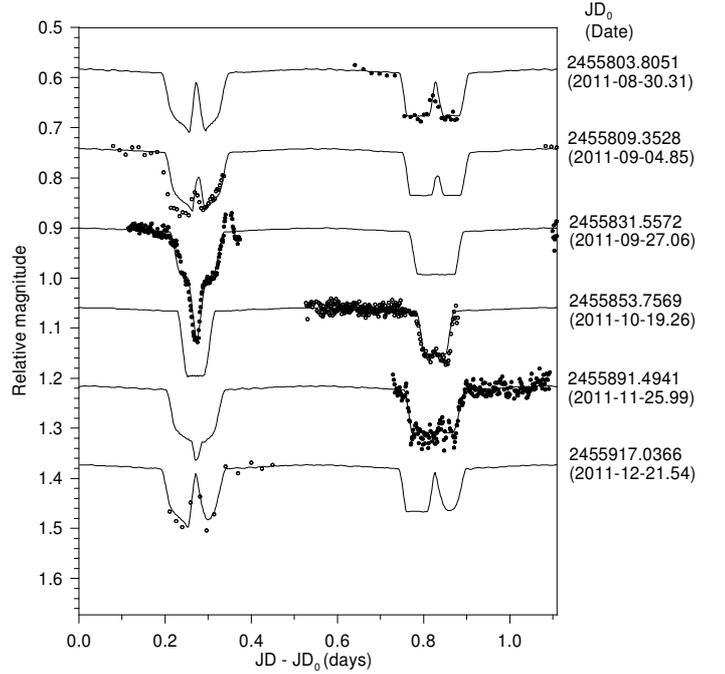}%
  \caption[Long-term evoluation of the orbital lightcurve of Isberga]{%
  \label{fig: lc2}
  Selected data of the long-period lightcurve component of Isberga during the
  2011 apparition, showing the long term evolution.
  The observed data are marked as points. The solid curve
  represents the synthetic lightcurve of the best-fit solution. 
  }
\end{center}
\end{figure}

    \indent We model the system using the method
    described in \citet{2009-Icarus-200-Scheirich},  
    modified to allow for precession of the orbit's pericenter.
    For the modeling, the optical lightcurves were reduced using the 
    technique described in \citet{2006-Icarus-181-Pravec}. In particular, 
    the rotation-induced lightcurve of the primary was fitted using 
    Fourier series and subtracted from the data. 
    The shapes of the components are modeled as ellipsoids,
    an oblate spheroid for the primary
    ($A_1$\,$=$\,$B_1$\,$>$\,$C_1$)
    and a prolate spheroid for the secondary
    ($A_2$\,$>$\,$B_2$\,$=$\,$C_2$),
    and approximated by polyhedra with triangular facets,
    orbiting each other on Keplerian orbits. 
    We assume \ben{same albedo and density for both components.
      This assumption is required to translate the unknown mass
      and diameter ratio of the components into flux ratio
      \citep[see][]{2009-Icarus-200-Scheirich}.
      Depending on the formation scenario, the satellite's density may however
      be different from that of the primary:
      under-dense for ejecta re-accumulation or over-dense for ejected
      boulder (unlikely here given the sizes of Isberga and its satellite).
    The} secondary is moreover assumed to be spin-orbit locked, its long
    axis aligned with the centers of the two bodies at the
    pericenter. 
    Finally, spin vectors of both components are assumed to be
    colinear with mutual orbit pole.\\ 
    \indent The total brightness of the system as seen by the observer was computed
    as a sum of contributions from all visible facets, using a ray-tracing 
    code that checks which facets are occulted by or in shadow from the other
    component. 
    In modeling the eccentric orbit, a precession of the line of apsides was
    taken into account. A pericenter drift rate depends on primary's 
%    polar flattening \citep[see][Eq. 6.249]{1999-Book-Murray}
    oblateness \citep[$A_1$/$C_1$, see][Eq. 6.249]{1999-Book-Murray}
    that is only poorly
    estimated from the lightcurves (see Table~\ref{tab: param}), so we fit the
    pericenter drift rate as an independent parameter
    \ben{($\dot{\omega}$)}.
    Its initial values were
    stepped in a range from zero to $30^{\circ}$/day; this range encompasses
    all possible values for the flattening and other parameters of the
    system. To reduce a complexity of the modeling, the upper limit on
    eccentricity is estimated by fitting data from the best-covered
    apparition (2011) only. \\
    \indent \ben{The fitted parameters are the oblateness of the primary, expressed
    as its equatorial-to-polar semi-major axes ratio, $A_1/C_1$; an elongation of the
    secondary, expressed as its equatorial (the largest) to polar (the
    shortest) semi-major axes ratio, $A_2/C_2$; a ratio between the
    mean cross-section equivalent diameters of
    the components of the binary ($D_{2,C}/D_{1,C}$); the pole coordinates of the mutual
    orbit in ecliptic frame, L$_{\rm p}$ and B$_{\rm p}$ (Epoch J2000); a
    relative size of the mutual orbit's semi-major axis ($a/D_{1,C}$);
    the mean length $L_0$ (\ie, the sum of angular distance from the
    ascending node and the length of the ascending node) for a given epoch
    $T_0$;
%    an argument
%    of mean length $L_0$ (\ie, the angular distance from the ascending node)
%    for a given epoch $T_0$;
    the sidereal orbit period $P_{\rm orb}$;
    and for modeling the eccentric orbit, the eccentricity e; and an
    argument of pericenter ($\omega$) as well. }\\ 
    \indent We obtain a unique prograde solution of the mutual orbit.
    The best-fit model parameters are given in
    Table~\ref{tab: param}, with uncertainties corresponding to
    3-$\sigma$ confidence level
    \citep[see][]{2009-Icarus-200-Scheirich}. 
    The orbital pole coordinates of the system,
    at a high ecliptic latitude (Fig.~\ref{fig: spin}),
    implies a small obliquity of $1.5^{+6.0}_{-1.5}$ deg. 
    Mutual events are therefore constantly observable from Earth,
    although the geometry remains limited to the equatorial region,
    \ben{precluding} a detailed modeling of the 3-D shape of the primary.
    \ben{We constrain the equatorial axes 
    ($A_1$ and $B_1$) from the amplitude of lightcurves at low phase
    angle and find $A_1$/$B_1$\,=\,1.23\,$\pm$\,0.02.
    The oblateness of the primary $A_1/C_1$ is, however,
    loosely constrained, with possible values ranging from 
    1.2\footnote{\ben{By definition of the ellipsoid,
      $A_1$\,$=$\,$B_1$\,$>$\,$C_1$, $A_1/C_1$ is thus larger or equal
        to $A_1$/$B_1$.}} to 2.0.}
    We do not see any evidence\footnote{\ben{The elongation of the
        secondary is indicated by the amplitude of the long-period
        component of the lightcurves outside mutual events, which is
        zero or very low here.}}  
    for a strong elongation of the satellite
    ($A_2$/$C_2$), even in the long lightcurve observations (6--8\,h)
    that cover a fourth of its rotation period (if it is indeed
    spin-orbit locked). 
    Examples of the data for the orbital lightcurve component together with
    the synthetic lightcurve for the best-fit solution 
    are presented in Fig.~\ref{fig: lc2}.

%%%%%%------ TABLE --- Begin --- Orbit ------%%%%%%
\begin{table}[t]
\begin{center}
%%%---Caption and Labeling ---%%%
  \caption[Isberga orbital and shape parameters]{%
    \ben{Best-fit values for a circular mutual
    orbit with 3-$\sigma$ uncertainties of the parameters described in
    Section~\ref{sec: orblc}.} 
    \label{tab: param}}
%%%--- Table Content ---%%%
  \begin{tabular}{ccc}
    \hline
    \hline
    Parameter & Value & Unit \\
    \hline
    (L$_{\rm p}$, B$_{\rm p}$) & (225, +86)$^a$            & deg. \\
    $P_{orb}$           & $26.6304 \pm 0.0001$   & h \\
    $L_0$              & 354 $\pm$ 3 & deg. \\
    $T_0$              & 2453790.631170   & JD\\ %(2006-02-24 03:08:53 UT)
    $e$                & $ \leq 0.10^b$          &\\
    $\omega$           & 0 -- 360 & deg. \\
    $\dot{\omega}$     & 0 -- 10 & deg. \\
    $a/D_{1,C}$         & $2.5^{+0.3}_{-0.6}$       &  \\
    $D_{2,C}/D_{1,C}$    &  $0.29 \pm 0.02$       & \\
    $P_{rot}$           & $2.91695 \pm 0.00010$   & h \\
    $A_1$/$C_1$         & $1.3^{+0.7}_{-0.07}$ &  \\
    $A_2/C_2$           & $1.1^c$                 &\\
    \hline
  \end{tabular}
\end{center}
{\small
%  {\it Note.} \\ 
  ${}^a$ The 3-$\sigma$ area is approximately an ellipse of semi-major
  axes of 8\degr~and 6\degr, centered on these coordinates, see
  Fig.~\ref{fig: spin}.\\ 
  ${}^b$ We estimated only an upper limit on the eccentricity from
  2011 data. \\
  ${}^c$ This is only a formal best-fit value of the elongation of the
  secondary, a spherical shape is consistent as well.
}
\end{table}
%%%%%%------ TABLE ---  End  --- Orbit ------%%%%%%

%%%%%%------ FIGURE --- Begin --- spin ------%%%%%%
%%%%%%------ FIGURE ---  End  --- spin ------%%%%%%

%%%%%%------ FIGURE --- Begin --- LC2 ------%%%%%%
%%%%%%------ FIGURE ---  End  --- LC2 ------%%%%%%

  \subsection{Interferometry analysis and size of the system\label{sec: orbmidi}}
    \indent To model and interpret the spatial information encoded in the correlated flux, we develop
    here an extension to the simple model of binary asteroids proposed by
    \citet{2009-ApJ-694-Delbo}, where the system was described by two
    uniform disks. Here, we model the primary component of the system by
    a uniform ellipse, thus taking into account the primary ellipsoid
    shape and rotation. We denote $\theta_{\alpha}$ and
    $\theta_{\beta}$ (with $\theta_{\alpha}>\theta_{\beta}$) the
    apparent major axes of the ellipse representing the primary
    component, and $\theta_2$ the apparent diameter of the secondary
    component. The two components are separated by the angular
    distance vector $\boldsymbol{\Lambda}$. The correlated flux
    ($F_{\rm B}$) produced by such a binary system is: 

{\footnotesize \begin{equation}
  \label{eq: flux}
  F_{\rm B}(\lambda) = F_1(\lambda,\theta_{\alpha},\theta_{\beta})
                          \left[\mathcal{V}_{1}^2(\lambda) + 
                                \mathcal{V}_{2}^2(\lambda)f_{21}^2 + 
                               2\mathcal{V}_{1}(\lambda)\mathcal{V}_{2}(\lambda)f_{21}\cos(\frac{2\pi\boldsymbol{B}}{\lambda}.\boldsymbol{\Lambda})
                          \right]^{\frac{1}{2}}
\end{equation}}

    \noindent where $F_1(\lambda,\theta_{\alpha},\theta_{\beta})$ is the total flux of the first
    component, 
%    $f_{21}$\,=\,$\left(\frac{\theta_2}{\sqrt{\theta_{\alpha}\theta_{\beta}}}\right)^2$
    $f_{21}$\,=\,$(\theta_2 / \sqrt{\theta_{\alpha}\theta_{\beta}})^2$
    is the flux ratio between
    the secondary and primary components, $\boldsymbol{B}$ is the baseline
    vector projected on the plane of the sky, and $\mathcal{V}_i$
    are the intrinsic normalized visibilities\footnote{\ben{The
        normalized visiblity is the ratio between the correlated flux
        and the total flux.}} of each component $i$, computed as 

    {\small \begin{equation}
        \mathcal{V}_i (\lambda) =  2 \frac{J_1(\pi \theta_i \frac{B}{\lambda})}{\pi \theta_i \frac{B}{\lambda}}
        \label{eq:ud}
    \end{equation}}
%    {\small \begin{eqnarray}
%        C_1 (\lambda)&=& 2 \frac{J_1(\pi \theta_{\alpha} \frac{B_{\rm eff}}{\lambda})}{\pi \theta_{\alpha} \frac{B_{\rm eff}}{\lambda}}\\
%        C_2 (\lambda)&=& 2 \frac{J_1(\pi \theta_2 \frac{B}{\lambda})}{\pi \theta_2 \frac{B}{\lambda}}
%    \end{eqnarray}}

    \noindent with $J_1$ the Bessel function of the first kind of
    order one, and $B$ the length of the projected baseline on the
    plane of the sky. However, a complexity has to be included in the
    derivation of the visibility expression for the primary component
    that we assumed to be elliptic and not circular anymore. An
    ellipse can be seen as a disk that is compressed along one of its
    axis, thus becoming the semi-minor axis, and then possibly rotated
    to define the semi-major axis orientation.
    Therefore, the link between the visibility of a circularly
    symmetric brightness distribution (uniform disk) and of its
    inclined and rotated version (ellipse), is obtained by a proper
    change in the baseline reference frame. This change takes into
    account a rotation followed by a compression factor along the
    proper baseline axis \citep[see][for more
      details]{2007-NewAR-51-Berger}. This leads to the concept of 
    \textsl{effective baseline}:
    \begin{equation}
      B_{\rm eff}=\sqrt{B_{u,\gamma}^2+B_{v,\gamma}^2\cos\left(\frac{\theta_\alpha}{\theta_\beta}\right)},
    \end{equation}

    \noindent which is the length of projected baseline expressed in the
    equatorial reference frame rotated by the angle $\gamma$, 
    the position angle, counted from
    North ($v$ coordinates) to East ($u$ coordinates), of the binary
    separation vector, with  
    {\small \begin{eqnarray}
        B_{u,\gamma}&=& B_{\rm u}\cos(\gamma)-B_{\rm v}\sin(\gamma)\\
        B_{v,\gamma}&=& B_{\rm u}\sin(\gamma)+B_{\rm v}\cos(\gamma)
    \end{eqnarray}}

    \indent In this rotated frame, the object recovers a circularly symmetric
    shape and the visibility expression of the elliptic brightness
    distribution is thus obtained by replacing, in Eq.\ref{eq:ud}, the projected baseline length $B$ (as
    defined in the initial equatorial reference frame) by the
    effective baseline length $B_{\rm eff}$, and $\theta_i$ by the
    longest axis $\theta_\alpha$. 
 
\begin{figure}[t]
\begin{center}
  \includegraphics[width=.49\textwidth]{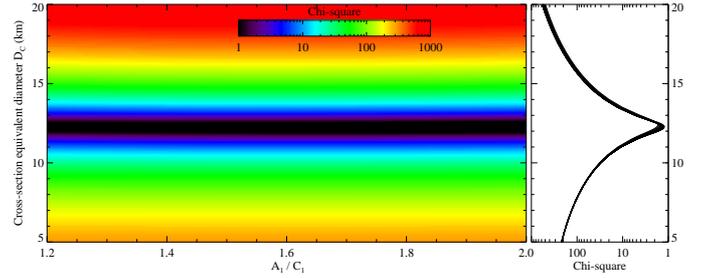}%
  \caption[Goodness of fit for the diameter determination]{%
  \label{fig: chi2}
  Goodness of fit for the determination of Isberga effective
  diameter and oblateness ($A_1$/$C_1$). }
\end{center}
\end{figure}

    \indent Owing to the low level of the correlated flux of Isberga, we
    analyze the four epochs together rather than independently, 
    tying all the parameters (4 for each epoch:
    $\theta_\alpha$, $\theta_\beta$, $\theta_2$, and $\Lambda$)
    to the effective, \ie, \ben{the cross-section equivalent,} diameter of 
    Isberga $D_C$. 
    We use the synthetic lightcurve $m(t)$ of the rotation
    component (see \ref{sec: orblc} and Fig.~\ref{fig: lc1}c) to express
    the apparent major axes of the primary component at each epoch,
    $\theta_\alpha(t)$ and $\theta_\beta(t)$, as a function of the
    \ben{cross-section} equivalent diameter $D_C$ and the primary oblateness
    $A_1$/$C_1$.
    First, assuming that the system is seen equator-on, which is a minor
    approximation as the latitude of the sub-Earth point is
    4\degr~only, $\theta_\beta(t)$ is constant and equals to the polar
    dimension $C_1$. 
    Second, the lightcurve amplitude provides the ratio of equatorial
    dimensions $A_1/B_1 = 10^{-0.4 ~ \left[ m_A - m_B \right] } = 1.23$, with
    $m_A$ and $m_B$ the minimum and maximum apparent magnitudes over a
    rotation.
    \ben{With these definitions, we write the \ben{cross-section} equivalent diameter as:}
    {\small \begin{equation}
    \nonumber
      \boldsymbol{D_C = \sqrt{<\theta_\alpha(t)\theta_\beta(t)>} = \sqrt{<\theta_\alpha(t)>C_1},} 	      
    \end{equation}}
    \ben{where $<.>$ is the temporal mean over one rotation of the primary. 
    Considering that $<\theta_\alpha(t)>=\sqrt{A_1B_1}$, we have:}
    {\small \begin{equation}
    \nonumber
      \boldsymbol{D_C = \sqrt{A_1 ~.~ 10^{0.2 ~ \left[ m_B - m_A \right]} C_1}=A_1\sqrt{\frac{10^{-0.2 ~ \left[ m_B - m_A \right]}}{A_1/C_1}}.} 	      
    \end{equation}}
    \ben{Using $\frac{\theta_\alpha(t)}{A_1}=10^{-0.4 ~ \left[ m(t) - m_A \right]}$, we find:}
    {\small \begin{equation}
      \theta_\alpha(t)= D_C ~.~ \sqrt{A_1/C_1} ~.~ 10^{0.1 ~ \left[ m_B - m_A \right] } ~.~ 10^{-0.4 ~ \left[ m(t) - m_A \right]}.
    \end{equation}}
    \indent With the total flux from the primary
    $F_1(\lambda,\theta_\alpha,\theta_\beta)$ evaluated using the NEATM
     \citep{1998-Icarus-131-Harris}, 
    the ratio between the component apparent diameters
    $f_{21}$ computed from their physical size ratio of 0.29 (see above,
    \ref{sec: orblc}), and the angular separation $\Lambda$ provided by
    the orbital solution, the free parameters are therefore restricted
    to the effective diameter $D_C$ and the oblateness $A_1$/$C_1$ of the
    primary (see Fig.~\ref{fig: midi} for a representation of the
    system configuration 
    at each epoch of observation).
    \ben{All other parameters are determined
    from these two free parameters. The $\eta$
      parameter is considered constant in our NEATM modeling
      \citep[using the value of $\eta$\,=\,1.211\,$\pm$\,0.022
        from][]{2011-ApJ-741-Masiero}}.\\  
    \indent We search for the best-fit solution
    by comparing the correlated flux of the model ($F_{\rm B,i}$),
    at each epoch $i$ and for each wavelength $j$, with the observations
    ($F_{i}^\prime$), for $D_C$ ranging from 5 to 20\,km and
    $A_1$/$C_1$ from 1.2 to 2.
    The goodness of fit indicator we use is 
    \begin{equation}  
      \chi^2 = 
        \sum_{i=1}^{N_{e}}
          \sum_{j=1}^{N_\lambda}
            \left(
              \frac{F_{\rm B,i}(\lambda_j) - F_{i}^\prime(\lambda_j)}{\sigma_{i,j}}
            \right )^2
      \label{eq: chi2}
    \end{equation}

    \noindent where $N_{e}$ is the number of epochs,
    $N_{\lambda}$ is the number of correlated flux samples at the epoch $i$,
    and $\sigma_{i,j}$ is the uncertainty on the measured correlated
    flux.
    \ben{We computed a grid of models by scanning $D_C$ between 5 and
      20\,km and $A_1$/$C_1$ between 1.0 and 2.0 (see
      Section~\ref{sec: orblc}). }\\      
    \indent As visible in Fig.~\ref{fig: chi2}, 
    the $\chi^2$ statistics is highly peaked around 12~km
    along the $D_C$ direction, while it is flatter along
    the $A_1$/$C_1$ direction. \ben{The best-fit to
    the data is thus obtained for a diameter 
    $D_C$\,=\,12.3\,$\pm$\,0.1\,km.} 
    Although we can not estimate accurately the quantitative
    contribution of the model systematics, we expect them to dominate
    the uncertainty budget and we adopt a more conservative value of 
    $D_C$\,$=$\,12.3\,$\pm$\,1.2\,km, \ie, a 10\% relative accuracy
    for diameter determination.
    \ben{The low contrast of the $\chi^2$ statistics does not convincingly  
      restricts the range of 1.2 to 2.0 derived from lightcurves
      (Sec.~\ref{sec: orblc}), although high oblateness seems to be
      favored by our modeling, with a formal best-fit value of
      $A_1/C_1=2.00^{+0.00}_{-0.45}$ derived 
      from a Bayesian analysis of the $\chi^2$ statistics.}\\
    \indent \ben{We present the best-fit model plotted together with
      the correlated flux and the 
    system geometry in Fig.~\ref{fig: midi}. Our best-fit solution is
    in best agreement with the fourth averaged measurement. This is
    expected given the smaller error bars and thus the stronger weight
    of this measurement in the fit process. Nevertheless, the best-fit
    model agrees with the other measurements within their error
    bars. We can however note a slight discrepancy around
    12-13~$\boldsymbol{\mu}$m for the first correlated flux
    measurement.}  \\
    \indent \ben{We use this diameter estimate and the absolute magnitude of
    12.18\,$\pm$\,0.27 we determine following the work by
    \citet{2012-Icarus-221-Pravec} and using observations with Trappist
    (Table~\ref{tab: lc})} to
    determine an albedo\footnote{\ben{We use the widely-used formula
        between the size $D$, the visible absolute magnitude $H$, and
        the geometric visible albedo $p_v$: $D({\rm
          km})=1329 p_v^{-1/2}10^{-H/5}$.}} of 0.14$^{+0.09}_{-0.06}$.
    \ben{S-type asteroids have higher albedo on average: 0.197\,$\pm$\,0.153
    \citep{2012-Icarus-221-Pravec}}. Such a 
    value is, however, within the range of possible albedo of S-types.\\
    \indent \citet{2011-ApJ-741-Masiero} reported a diameter of 
    $D_W$\,=\,10.994\,$\pm$\,0.067 km and an albedo of 
    0.21\,$\pm$\,0.02 based on a NEATM
    \citep{1998-Icarus-131-Harris} analysis of WISE mid-infrared
    radiometry. Taking into account the 
    binarity of Isberga, this converts into a \ben{cross-section} equivalent
    diameter for the primary of 10.5\,$\pm$\,0.1\,km, to be compared
    with our determination. Considering a 5--10\% uncertainty on the
    diameter determined from WISE to encompass possible systematics
    \citep[see the comparison of diameter estimates from thermal
      modeling with other methods in][]{2006-Icarus-185-Marchis,
      2012-PSS-73-Carry}, both determinations converge to a
    $\approx$11\,km surface-equivalent diameter for Isberga. 
    The smaller albedo determined here is due to the larger
    diameter determination.

%WISE: d=10.9940 +/- 0.06700
%     pv= 0.2058 +/- 0.01480
%  Deq sys = (1+0.29^2) power 1/2
%  DWISE/fac = 10.5
%wise compte prim + sat dans ses 11km
%

%%%%%%------ FIGURE --- Begin --- Chi2 ------%%%%%%
%%%%%%------ FIGURE ---  End  --- Chi2 ------%%%%%%

  \subsection{Physical properties of Isberga}

    \indent \ben{The results above restrict the primary oblateness
      $A_1$/$C_1$ between 1.2 and 2.0.
    Because the density determination from the lightcurve analysis is
    strongly dependent on this parameter, we fix the
    oblateness between 1.2 and 2.0 by steps of 0.1 and we determine 
    different density values ranging from
    $1.89^{+0.85}_{-1.00}$
    to
    $4.03^{+0.60}_{-1.99}$
    g.cm$^{-3}$ (3-$\sigma$ confidence interval).
    Since the ellipsoidal shape
    approximation tends to overestimate the volumes of the components,
    the derived bulk densities should be considered as lower limit estimates.
    Formally, the 3-$\sigma$ range for the density is therefore
%    $3.44^{+2.24}_{-1.18}$ g.cm$^{-3}$.
    $2.91^{+1.72}_{-2.01}$ g.cm$^{-3}$.
    This level of accuracy corresponds to about 40\% relative accuracy at
    1-$\sigma$ level. This crude precision is, however, better than
    that of 45\% of all density determinations \citep[see Fig.~3
      in][]{2012-PSS-73-Carry}. This highlights
    the yet limited knowledge on asteroid interiors.} \\
    \indent \ben{This density of 2.91$^{+1.72}_{-2.01}$\,g.cm$^{-3}$
      is very close to 
      the typical density of S-type 
      asteroids at 2.72\,$\pm$\,0.54\,g.cm$^{-3}$
      \citep[][]{2012-PSS-73-Carry}.
      This density is lower than the grain density of the
      associated H ordinary chondrite meteorites of 
      3.72\,$\pm$\,0.12\,g.cm$^{-3}$
      \citep{2008-ChEG-68-Consolmagno}.
      The porosity of Isberga is therefore 
      22$^{+54}_{-22}$\%, and its macroporosity can be estimated to
      14$^{+66}_{-14}$\%
      \citep[using a microporosity of 7.0\,$\pm$\,4.9\% on H
        chondrites measured by][]{2008-ChEG-68-Consolmagno}.}\\
    \indent \ben{The internal structure of Isberga thus encompasses all
      possible, from compact to highly porous. Although the presence
      of some macroporosity is likely, better constrains
      on Isberga polar oblateness are required to conclude. 
      From the current census of S-type densities, and the linear
      trend of asteroids to range from large and dense 
      to small and porous \citep[see Fig. 9 in][]{2012-PSS-73-Carry}, 
      it is, however, unlikely that Isberga has a density above
      $\approx$3\,g.cm$^{-3}$. We therefore favor solution with
      oblateness below 1.5--1.6.} \\
    \indent \ben{We finally use this density determination to estimate
      the mass of Isberga and of its satellite: from the primary
    volume-equivalent diameter
    $D_V = \left(\frac{A_1}{C_1}\right)^{-1/6} \left(\frac{A_1}{B_1}\right)^{-1/12} D_C$
    (Table \ref{tab: summary}), we find
    $M_1$\,=\,$3.52^{+3.90}_{-2.73}$ $\times$ 10$^{15}$\,kg
    and
    $M_2$\,=\,$8.60^{+24.1}_{-7.83}$ $\times$ 10$^{13}$\,kg, 
    respectively. 
    The size of the Hill sphere associated with these masses is of
    2320$^{+650}_{-700}$ and 670$^{+330}_{-370}$\,km. The system is
    therefore extremely compact, the 
    components being separated by only 33\,km.}

%
%%%%%%%------ TABLE --- Begin --- Isberga -- Summary ------%%%%%%
\begin{table}[t]
\begin{center}
%%%---Caption and Labeling ---%%%
  \caption[Physical characteristics of the binary (939) Isberga]{
    \label{tab: summary}
    Physical characteristics of the binary (939) Isberga.  
    We list the characteristic of the mutual orbit and 
    for both the primary and the satellite their surface- and
    volume-equivalent diameter, density, and mass. For the primary, we
    also report the axes ratios and rotation period.
    \ben{Uncertainties are 3-$\sigma$.
      Values of A$_2$/B$_2$ and A$_2$/C$_2$ are formal
      best-fit, but values 1.0 are consistent with the data as well.}
}
%%%--- Table Content ---%%%
  \begin{tabular}{cccc}
    \hline\hline
    Parameter & Value & Unit \\
    \hline
    \noalign{\smallskip}
    \multicolumn{2}{l}{\hspace{0em}Primary}\\
%    \noalign{\smallskip}
    $D_{1,C}$     & 12.3 $\pm$ 1.2                         & km \\
    $D_{1,V}$     & 12.4$^{+2.5}_{-1.2}$                   & km \\
    $\rho_1$     &  2.91$^{+1.72}_{-2.01}$                  & g.cm$^{-3}$ \\
    $M_1$        & $3.52^{+3.90}_{-2.73}$ $\times$ 10$^{15}$ & kg \\
    A$_1$/B$_1$   & 1.23  $\pm$ 0.02 \\
    A$_1$/C$_1$   & 1.3$^{+0.7}_{-0.03}$  \\
    $P_{rot}$      & $2.91695 \pm 0.00010$   & h \\
    \noalign{\smallskip}
    \hline
    \noalign{\smallskip}
    \multicolumn{2}{l}{\hspace{0em}Satellite}\\
%    \noalign{\smallskip}
    $D_{2,C}$     & 3.6 $\pm$ 0.5                          & km \\
    $D_{2,V}$     & 3.6$^{+0.7}_{-0.3}$                      & km \\
    $\rho_2 \equiv \rho_1$ &  2.91$^{+1.72}_{-2.01}$        & g.cm$^{-3}$ \\
    $M_2$        & $8.60^{+24.1}_{-7.83}$ $\times$ 10$^{13}$ & kg \\
    A$_2$/B$_2$   & 1.1  \\
    A$_2$/C$_2$   & 1.1  \\
    $P_{rot} \equiv P_{orb}$      & $26.6304 \pm 0.0001$   & h \\
    \noalign{\smallskip}
    \hline
    \noalign{\smallskip}
    \multicolumn{2}{l}{\hspace{0em}Mutual orbit}\\
%    \noalign{\smallskip}
    $a$       & $33.0^{+7.6}_{-1.4}$       &  km \\
    ($\lambda_{\rm p}$, $\beta_{\rm p}$) & (225, +86) $\pm$ 7    & deg. \\
    $P_{orb}$          & $26.6304 \pm 0.0001$   & h \\
    $e$               & $ \leq 0.10$          &\\
    \noalign{\smallskip}
    \hline
  \end{tabular}
\end{center}
\end{table}
%%%%%%%------ TABLE ---  End  --- Isberga -- Summary ------%%%%%%

%%%%%%%%%%%%%%%%%%%%%%%%%%%%%%%%%%%%%%%%%%%%%%%%%%%%%%%%%%%%%%%%%%%%%%%%%%%%%
%%% TAG %%%------%%%%%%%%%%%%%%%%%%%%%%%%%%%%%%%%%%%%%%%%%%%%%%%%%%%%%%%%%%%%
%%%%%%%%%%%%%%%%%%%%%%%%%%%%%%%%%%%%%%%%%%%%%%%%%%%%%%%%%%%%%%%%%%%%%%%%%%%%%
\section{Conclusion}
  \indent We present the first mid-infrared interferometric
  observations of a small binary asteroid, (939) Isberga. Together with
  low-resolution spectroscopy in the near infrared and an extensive
  campaign of lightcurves, we conduct a complete characterization of
  the surface, orbital, and physical properties of the system.
  It is composed by a 13\,km S-type primary and a 4\,km secondary, orbiting
  each other in 26\,h on a \ben{nearly}-circular orbit of semi-major axis
  33\,km, deep inside the Hill sphere.
  The inferred density of \ben{$2.91^{+1.72}_{-2.01}$ g.cm$^{-3}$ is typical
  for this composition, but the large uncertainties prevent from
  concluding on the internal structure.}
  The system has a low obliquity ($1.5^{+6.0}_{-1.5}$ deg.) and mutual
  eclipses and occultations are always visible from Earth. More
  lightcurve observations of the system, in particular with absolute
  photometric reference, will help constrain better the elongation of
  the secondary and the eccentricity of the mutual orbit. \\
  \indent The low mid-infrared flux of Isberga, at the very edge of
  VLTI/MIDI capabilities, \ben{precludes} an independent analysis of these
  data. The combined analysis of optical lightcurves and mid-infrared
  visibilities we present here is, however, an efficient way to 
  derive both relative quantities and absolutely scale the system.
  Among the many small main-belt binaries, all present
  similar mid-infrared fluxes (given their temperature and apparent
  angular size), and more sensitive instruments/modes, like the
  forthcoming VLTI/MATISSE, must be used to
  better characterize this population of binary systems.

\section*{Acknowledgments}

  \indent We acknowledge support from the Faculty of the European
  Space Astronomy Centre (ESAC) for the visits of
  M. Delbo and A. ``Momo'' Matter.
  The work by P.S. and P.P. was supported by the Grant Agency of the
  Czech Republic, Grant P209/12/0229, and by Program RVO 67985815.
  A. Matter acknowledges financial support from the Centre National
  d'{\'E}tudes Spatiales (CNES).
  \ben{TRAPPIST is a project funded by the Belgian Fund for
    Scientific Research (Fonds National de la Recherche Scientifique,
    F.R.S.-FNRS) 
    E. Jehin and M. Gillon are FNRS Research Associates, and 
    Jean Manfroid is Research Director of the FNRS.}
  Part of the data utilized in this publication were obtained and made
  available by the The MIT-UH-IRTF Joint Campaign for NEO Reconnaissance. The
  IRTF is operated by the University of Hawaii under Cooperative Agreement
  no. NCC 5-538 with the National Aeronautics and Space Administration, Office
  of Space Science, Planetary Astronomy Program. The MIT component of this
  work is supported by NASA grant 09-NEOO009-0001, and by the National Science
  Foundation under Grants Nos. 0506716 and 0907766. 
  FED acknowledges funding from NASA under grant number NNX12AL26G and
  Hubble Fellowship grant HST-HF-51319.01-A. 
  Any opinions, findings, and conclusions or recommendations expressed in this
  material are those of the authors and do not necessarily reflect the views
  of NASA or the National Science Foundation.
  \ben{This research utilizes spectra acquired by Jeffery F. Bell
    with the NASA RELAB facility at Brown University}

%% References with bibTeX database:

%\bibliographystyle{elsarticle-harv}
%\bibliography{biblio}

%% Authors are advised to submit their bibtex database files. They are
%% requested to list a bibtex style file in the manuscript if they do
%% not want to use elsarticle-harv.bst.

\end{document}